\documentclass[12pt]{article}
\usepackage[dvips]{graphicx}
\newcommand{\diag}{\mathop{\rm diag}\nolimits}
\newcommand{\tr}{\mathop{\rm tr}\nolimits}
\newcommand{\bz}{{\bf z}}
\newcommand{\dbz}{\overline{\bz}}
\newcommand{\hbz}{\hat{\bz}}
\newcommand{\bbz}{\breve{\bz}}
\newcommand{\bH}{{\bf H}}
\newcommand{\bbH}{\breve{{\bf H}}}
\newcommand{\bJ}{{\bf J}}
\newcommand{\bQ}{{\bf Q}}
\newcommand{\bW}{{\bf W}}
\newcommand{\bbW}{\breve{\bf W}}
\newcommand{\bK}{{\bf K}}
\newcommand{\bM}{{\bf M}}
\newcommand{\bbM}{\breve{\bf M}}
\newcommand{\bN}{{\bf N}}
\newcommand{\bE}{{\bf E}}
\newcommand{\bR}{{\bf R}}
\newcommand{\bU}{{\bf U}}
\newcommand{\dbU}{\overline{\bU}}
\newcommand{\bbU}{\breve{\bU}}
\newcommand{\dd}{\overline{d}}
\newcommand{\dq}{\overline{q}}
\newcommand{\bq}{\breve{q}}
\newcommand{\hq}{\hat{q}}
\newcommand{\dx}{\overline{x}}
\newcommand{\ddp}{\overline{p}}
\newcommand{\bp}{\breve{p}}
\newcommand{\hp}{\hat{p}}
\newcommand{\brho}{\breve{\rho}}
\title{On the circular Sitnikov problem: the alternation of stability and instability in
the family of vertical motions}
\author{Vladislav V. Sidorenko \\
{\footnotesize Keldysh Institute of Applied Mathematics} \\
{\footnotesize Russian Academy of Sciences } \\
{\footnotesize Miusskaya sq., 4., 125047 Moscow (Russia)}\\
{\footnotesize \texttt{(sidorenk@spp.keldysh.ru)}} }
\date{}
\begin{document}
\maketitle
\begin{abstract}

This paper is devoted  to the special case of the restricted
circular three-body problem, when the two primaries are of equal
mass, while the third body of negligible mass performs
oscillations along a straight line perpendicular to the plane of
the primaries (so called periodic vertical motions). The main
goal of the paper is to study the stability of these periodic
motions in the linear approximation. A special attention is given
to the alternation of stability and instability within the family
of periodic vertical motions, whenever their amplitude is varied
in a continuous monotone manner.
\end{abstract}

\section{Introduction}

The term ``Sitnikov problem" appeared originally in the context of
studies of oscillatory solutions in the restricted three body
problem. These studies were initiated by Sitnikov~\cite{S1960};
they stimulated the application of symbolic dynamics in celestial
mechanics~\cite{A1968}. We recall that Sitnikov considered the
case when two primaries have equal masses and rotate around
their barycenter $O$, while the infinitesimal third
body moves along a straight line normal to the plane defined by
the motion of the primaries and passing through $O$ (usually the
motions of the third body perpendicularly to the plane of the
primaries are called ``vertical"; below we will follow this
tradition).

Sitnikov concentrated his attention on phenomena taking place
when the primaries move in elliptic orbits. More bibliography on ``elliptic''
Sitnikov problem can be found, for example, in~\cite{H2009,HL2005,KE2009}.

If the primaries move in circular orbits, then the vertical motions are integrable. The
corresponding quadratures were presented at the beginning of the
XX century by Pavanini~\cite{P1907} and MacMillan~\cite{MM1911} -
much before the start of Sitnikov's studies. Relatively simple
formulae for the vertical motions, written in terms of Jacobi
elliptic functions, can be found in~\cite{BLO1994}.

Since the integrability of third body motion is something
extraordinary within the restricted three body problem, many
specialists investigated the properties of vertical motions in
the case of primaries moving on circular orbit. Very often the
term ``circular Sitnikov problem" is applied to describe this
field of research. Taking into account its popularity, we will use
it too. Nevertheless, some authors prefer terms like ``Pavanini
problem" or ``MacMillan problem", which are probably more correct
from the historical point of view.

Depending on the initial values, three types of vertical motions
are possible in the circular Sitnikov problem: the hyperbolic
escape (i.e., the escape of the third body with non-zero velocity
at infinity), the parabolic escape (i.e., the escape of the third
body with zero velocity as the limit at infinity) and, finally,
the periodic motion, in which third body goes away up to a
distance $a$ from the plane defined by primaries and then returns
to it.

The first stability analysis of the periodic vertical motions in
the circular Sitnikov problem was undertaken by Perdios and
Markellos~\cite{PM1988}, but they drew the wrong conclusion that
vertical motions are always unstable (Perdios and
Markellos only analyzed the vertical motions with the initial conditions
such that $a<4$; as it was established lately it is not enough
to put any hypothesis about the stability properties of the motions
with larger values of $a$). The mistake was pointed
out in~\cite{BLO1994}, where the alternation of stability and
instability of vertical motions were found numerically in the
case of continuous monotone variation of their amplitude $a$.
Lately the existence of such an alternation was confirmed by the
results of computations presented in~\cite{P2007} and
\cite{SBD2007}. Taking into account their numerical results, the
authors of~\cite{SBD2007} proposed the hypothesis that the
lengths of stability and instability intervals have finite limits
as $a$ increases. This hypothesis was formulated on the basis of
computations in which $a$ did not exceed the value $13$. Our
numerical investigations demonstrate that the rapidly decreasing
difference of the stability intervals at $a\approx 13$ is a
manifestation of a local maximum of their lengths; if $a$ is
increased further, then the lengths of the stability and
instability intervals tend to zero.

There is one more important property of vertical motions, which
can be observed only for $a\gg 1$: the intervals of ``complex
saddle" instability, when all eigenvalues of the monodromy matrix
are complex and do not lie on the unit circle. According to our
computations first such an interval begins at $a \approx 546.02624$,
its length is $\approx 10^{-5}$. It means the erroneous of the statement
in~\cite{BLO1994} (p. 113), that the stability indexes of the vertical motions in
circular Sitnikov problem are always real (this statement was based on the results
of numerical studies in which the amplitude of the motion $a$ was smaller $17$; as
one can see it was not enough for such a general conclusion).

To conclude our short review on previous investigations of
vertical motions' stability in circular Sitnikov problem we would
like to mention the generalization of this problem for systems of
four and more bodies~\cite{BP2009, SPB2008}. Numerical results
presented in ~\cite{BP2009, SPB2008} demonstrate that in the
generalized problem the absence of stability/instability
alternation in the family of vertical motions persists.

The aim of our paper is to study the stability property of the
periodic vertical motions at large values of the ``oscillation
amplitude" $a$, both numerically and analytically. A special
attention will be given to the phenomenon of infinite alternation
of stability and instability in this family.

In fact, the infinite alternation of stability and instability in
the one-parameter family of periodic solutions is rather typical
for Hamiltonian systems, although the general investigation was
carried out only for 2DOF systems~\cite{CPR1980, GR1997}.
Different examples can be found in~\cite{CZ1983, H1983, NS2000}.

Nevertheless, an important difference exists between the circular
Sitnikov problem and other systems in which the alternation of
stability and instability was established earlier. In the circular
Sitnikov problem the discussed family of periodic solutions
possesses as a limit unbounded aperiodic motions - parabolic
escapes, while in previously considered systems the corresponding
families and their aperiodic limits were bounded~\cite{CZ1983,
NS2000}. Due to this difference, the alternation of stability and
instability in the circular Sitnikov problem can not be studied in
the same way as it was done in~\cite{CZ1983, H1983, NS2000} (one
could try to compactify the phase space by means of certain
changes of variables, but we were unable to find any reduction to
what was investigated already).

This paper is organized as follows. In Sect. 2 some general
properties of the vertical motions are discussed. In Sect. 3 we
present the linearized motion equations used in our studies of
the vertical motions' stability. The results of the numerical
investigation of the stability are reported in Sect. 4. In Sect.
5 we prepare for the analytical investigation: the approximate
expression for the monodromy matrix is derived here. Using this
expression, some important stability properties of vertical
periodic solutions with large amplitudes $a$ are established in
Sect. 6 (in particular, the asymptotic formulae for the intervals
of stability and instability are obtained). In Sect. 7 we discuss
briefly the vertical motions in the generalized circular Sitnikov
problem with four and more bodies. Some concluding remarks can be
found in Sect. 8.

\section{Preliminary. Some general properties of the vertical motions
in the circular Sitnikov problem}

We consider the restricted, circular, three-body problem with
primaries having equal masses, say $m_1 = m_2 = m$. Let $O x_1 x_2
x_3$ be a synodic (rotating) reference frame with the origin at
the barycenter $O$; the masses $m_1$ and $m_2$ are arranged on the
axis $Ox_1$, while the axis $Ox_3$ is directed along the rotation
axis of the system. The coordinates of the infinitesimal third
body in the synodic reference frame will be used as generalized
variables:
$$
q_1 = x_1,\quad q_2 = x_2,\quad q_3 = x_3.
$$
Below we assume that all variables are dimensionless.

The equations of motion of the third body can be written in
Hamiltonian form with Hamiltonian function~\cite{BLO1994}
$$
{\cal H}=\frac{1}{2}\left(p_1^2+p_2^2+p_3^2\right)+p_1q_2 -
p_2q_1 - \frac{1}{2}\left(\frac{1}{r_1}+\frac{1}{r_2}\right).
$$
Here $r_1$ and $r_2$ denote the distance between the third body
and the corresponding primary, while $p_1,p_2,p_3$ are the
momenta conjugated to $q_1,q_2,q_3$.

The phase space ${\cal V}=\{(p,q)\}$ possesses a manifold
$$
\tilde{\cal V}=\left\{(p,q),p_1=p_2=q_1=q_2=0\right\}
$$
which is invariant with respect to the phase flow. The phase
trajectories lying on $\tilde{\cal V}$ correspond to vertical
motions with the third body staing always on the axis $Ox_3$.
Consequently, the vertical motions are governed by a reduced 1DOF
system with Hamiltonian
$$
\tilde{\cal H}=\frac{p_3^2}{2}-
\frac{1}{\sqrt{q_3^2+\frac{1}{4}}}. \eqno(1)
$$
The phase portrait of the system with the Hamiltonian (1) is
shown in Fig. 1. It is remarkable that the separatrices (the
borders between trajectories representing periodic motions and
hyperbolic escapes) intersect at infinity.

\begin{figure}
\vglue-1.0cm \hglue2.cm
\includegraphics[width=12.0cm,keepaspectratio]{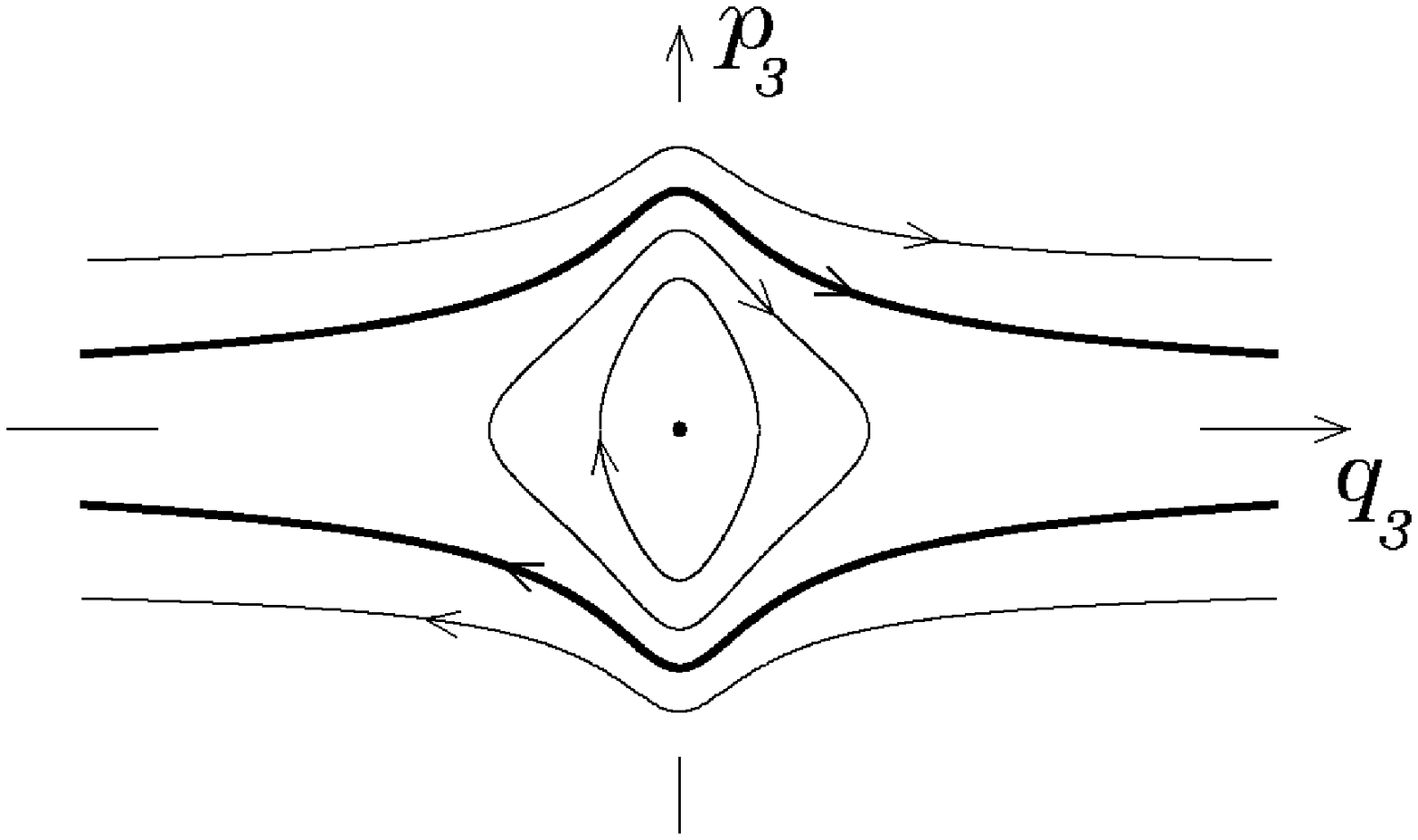}
\vglue-10cm \caption{Phase flow on the manifold $\tilde{\cal V}$.
Thick lines denote the separatrices ($\tilde{\cal H}=0$). }
\label{PP}
\end{figure}

The periodic solutions associated to the system with Hamiltonian
$\tilde{\cal H}$ form a one-parameter family
$$
p_3(t,a),q_3(t,a), \eqno(2)
$$
where as parameter $a$ one can choose the ``amplitude" of the
periodic motion (i.e., $a=\max_{t\in R^1}|\,q_3|$) or the
absolute value of $p_3$ at the passage trough the barycenter $O$
or the value $\tilde h$ of the Hamiltonian $\tilde{\cal H}$ in
this periodic motion. The first variant is the most convenient
for us, therefore $a$ in (2) will denote the ``amplitude" of the
periodic motion. For definiteness we assume that
$$
p_3(0,a)=0,\quad q_3(0,a)=a.
$$

There exist explicit expressions for the solutions (2) in terms of
Jacobi elliptic functions~\cite{BLO1994}. Since they are not used
in the forthcoming analysis, we do not rewrite them here, except
for the formula about the period of vertical motion:
$$
T=\frac{\sqrt{2}}{1-2k^2}\left[
E(k)+\frac{\pi}{2\sqrt{2(1-2k^2)}}\left( 1 - \Lambda_0\left(
\arcsin \sqrt{\frac{1-2k^2}{1-k^2}},k\right)\right)\right].
\eqno(3)
$$
Here $E(k)$ is the complete elliptic integral of the second kind,
$\Lambda_0(\varphi,k)$ is the Heuman Lambda Function, while the
value of the modulus $k$ is given by the formula
$$
k=\frac{1}{2}\sqrt{2+\tilde{h}},
$$
where
$$
\tilde{h}=-\frac{1}{\sqrt{a^2+\frac{1}{4}}}.
$$

For motions with large amplitudes ($a \gg 1$) the following
approximate formula can be used in place of (3):
$$
T\approx \sqrt{2}\pi a^{3/2}. \eqno(4)
$$

As it was mentioned before, the separatrices
$S^\pm=\{(p_3^{\pm}(t),q_3^{\pm}(t)),t\in R^1\}$, representing the
parabolic escapes, can be interpreted as a formal limit for
periodic motions at $a \to \infty$. The parabolic escapes obey
the approximate law
$$
q_3^{\pm}(t)\approx \pm \left(\frac{3}{\sqrt{2}}\right)^{2/3}
t^{2/3}. \eqno(5)
$$
Formulae (4) and (5) are easily obtained if one suitably relates
the properties of vertical motions with the properties of
rectilinear motions of a particle in a Newtonian field.

\section{The stability problem for periodic vertical motions}

Our efforts are concentrated on the analysis of the vertical
motions' stability with respect to ``horizontal" perturbations, due
to which the third body leaves the axis $Ox_3$. Under the linear
approximation, the behavior of the variables $p_1,p_2,q_1,q_2$ in
the perturbed motion is described by the linear Hamiltonian system
of equations with periodic coefficients:
$$
\frac{d\bz}{dt}=\bJ \bH(t)\bz. \eqno(6)
$$
Here
$$
\bz=(p_1,p_2,q_1,q_2)^T,
$$
$$
\bJ=\left(\begin{array}{cc}
{\bf 0}&-{\bf E}_2\\
{\bf E}_2&{\bf 0}\end{array}\right),\quad
\bH(t)=\left(\begin{array}{cccc} 1&0&0&1\\
0 & 1 & -1 & 0 \\
0 & -1 & \left(\frac{\displaystyle 1}{\displaystyle D^3}-
\frac{\displaystyle 3}{\displaystyle 4D^5}\right)& 0 \\
1 & 0 & 0 & \frac{\displaystyle 1}{\displaystyle D^3} \end{array}
\right).
$$
The symbol $\bE_k$ is used to denote the identity matrix of the
$k$-th order. The function
$D(t,a)=\left(q_3^2(t,a)+\frac{1}{4}\right)^{1/2}$ depends
periodically on time with a period $T_*=\frac{T(a)}{2}$, where
$T(a)$ denotes the period of the particular vertical motion whose
stability is investigated.

As it is known, the restricted circular three-body problem admits
several types of symmetry (for example, they are used for the
numerical construction of 3D periodic solutions~\cite{P2007}).
The consequence of these symmetries is the following property of
the variational equations (6): if $\bz(t)$ is a solution of (6),
then these equations admit the solution
$$
\tilde{\bz}(t)=\bQ\bz(-t), \eqno(7)
$$
where $\bQ$ it the $4\times4$-diagonal matrix,
$\bQ=\diag(1,-1,-1,1)$.

According to Floquet theory, in order to draw a conclusion about
the stability or instability of the solutions of (6), one should
analyze the spectral properties of the monodromy matrix
$\bM=\bW(T_*,0)$, where $\bW(t,t')$ denotes the normal
fundamental matrix corresponding to the system (6) (i.e., the
matrix solution of (6) with the initial condition
$\bW(t',t')=\bE_4$).

The normal fundamental matrix corresponding to the linear
Hamiltonian system (6) is a symplectic one, i.e.
$$
\bW^T(t,t')\bJ\bW(t,t')=\bJ.
$$
It is also worthwhile to mention some other properties of this
matrix:
$$
\bW(t,t'')=\bW(t,t')\bW(t',t''),\quad
\bW(t+T_*,t'+T_*)=\bW(t,t'), \eqno(8)
$$
$$
\bW(0,-t)=\bQ\bW^{-1}(t,0)\bQ=-\bQ\bJ\bW^T(t,0)\bJ\bQ.
$$
The first two equalities in (8) are elementary, while the last one
is a consequence of the symmetry property (7).

Using the relation (8) one easily obtains
$$
\bM=\bQ\bW^{-1}\left(\frac{T_*}{2},0\right)\bQ\bW\left(\frac{T_*}{2},0\right)=
-\bQ\bJ\bW^T\left(\frac{T_*}{2},0\right)\bJ\bQ\bW\left(\frac{T_*}{2},0\right).
$$

The characteristic equation of the system (6)
$$
\det\left(\bM - \rho \bE_4\right)=0 \eqno(9)
$$
is reciprocal and it can be written as
$$
\rho^4 - c_1 \rho^3 + c_2 \rho^2 - c_1 \rho + 1=0\,
$$
where
$$
c_1=\tr \bM\ ,\qquad c_2=\sum_{j=1}^3 \sum_{k=j+1}^4
(m_{jj}m_{kk}- m_{jk}m_{kj})\,.
$$
The quantities $m_{ij}$ in the last formula are the elements of
the monodromy matrix $\bM$.

It is also possible to rewrite the characteristic equation (9) as
the product
$$
(\rho^2 - 2b_1 \rho + 1)(\rho^2 - 2b_2 \rho + 1)=0\,. \eqno(10)
$$
The coefficients $b_1,b_2$ in (10) are the roots (real or
complex) of the quadratic equation:
$$
4x^2 - 2c_1 x + (c_2 - 2)=0.
$$
Often enough the quantities $b_1,b_2$ are called the stability
indices~\cite{BP2009}. The periodic vertical motion is stable
whenever $\{b_1,b_2\}\subset I=(-1,1)\subset R^1$ (i.e., when
$b_1,b_2$ are real and their absolute values are smaller than 1).
In the case
$$
\{b_1,b_2\}\subset \bar I=[-1,1],\quad
\{b_1,b_2\}\lefteqn{\subset}{\;\,/} \; I
$$
an additional investigation is needed to draw a conclusion about
stability or instability. In all other cases the instability
takes place.

\section{Numerical results}

We recall that in~\cite{BLO1994}  the alternation of the stability
and instability in the family of periodic vertical motions (2)
was discovered. Later on, more accurate results were published
in~\cite{SBD2007}: the length of the first 35 intervals of
stability and of the first 34 intervals of instability was
calculated. In~\cite{SBD2007} also an attempt was undertaken to
establish certain regularity in the variation of these
quantities: the existence of non-zero limits for the intervals'
lengths was proposed.

In Fig.2 and Fig.3 we present the results of some calculations,
when the first 700 intervals of stability and instability are
considered. The graph in Fig. 2 shows that for the first 30
intervals of stability the length of the intervals increases and
only afterwards the decrease of the length takes place. The
hypothesis formulated in~\cite{SBD2007} was based on the wrong
interpretation of the small variation of the intervals length in
vicinity of the maximum. In Fig. 3 the length of the instability
intervals decreases monotonically and it does not follow the
empirical law derived in~\cite{SBD2007} (according to this law,
the length of the instability intervals has the limit
$\Delta_{inst}\approx 0.254$; evidently, it is not so).

Our results allow us to propose the following approximate
formulae to characterize the behavior of the stability and
instability intervals' length in Fig. 2 and Fig. 3:
$$
\Delta_{st}\approx 0.25N^{-1/3},\quad \Delta_{inst}\approx
0.584N^{-1/3}. \eqno(11)
$$
More precisely, these formulae are valid for the periodic vertical
motions with amplitude $a$ smaller the critical value $a_*= 546.02624...$
The reason of such a restriction
and the situation for $a>a_*$ will be revealed a little bit later.

\begin{figure}
\vglue-0.0cm \hglue2.cm
\includegraphics[width=12.0cm,keepaspectratio]{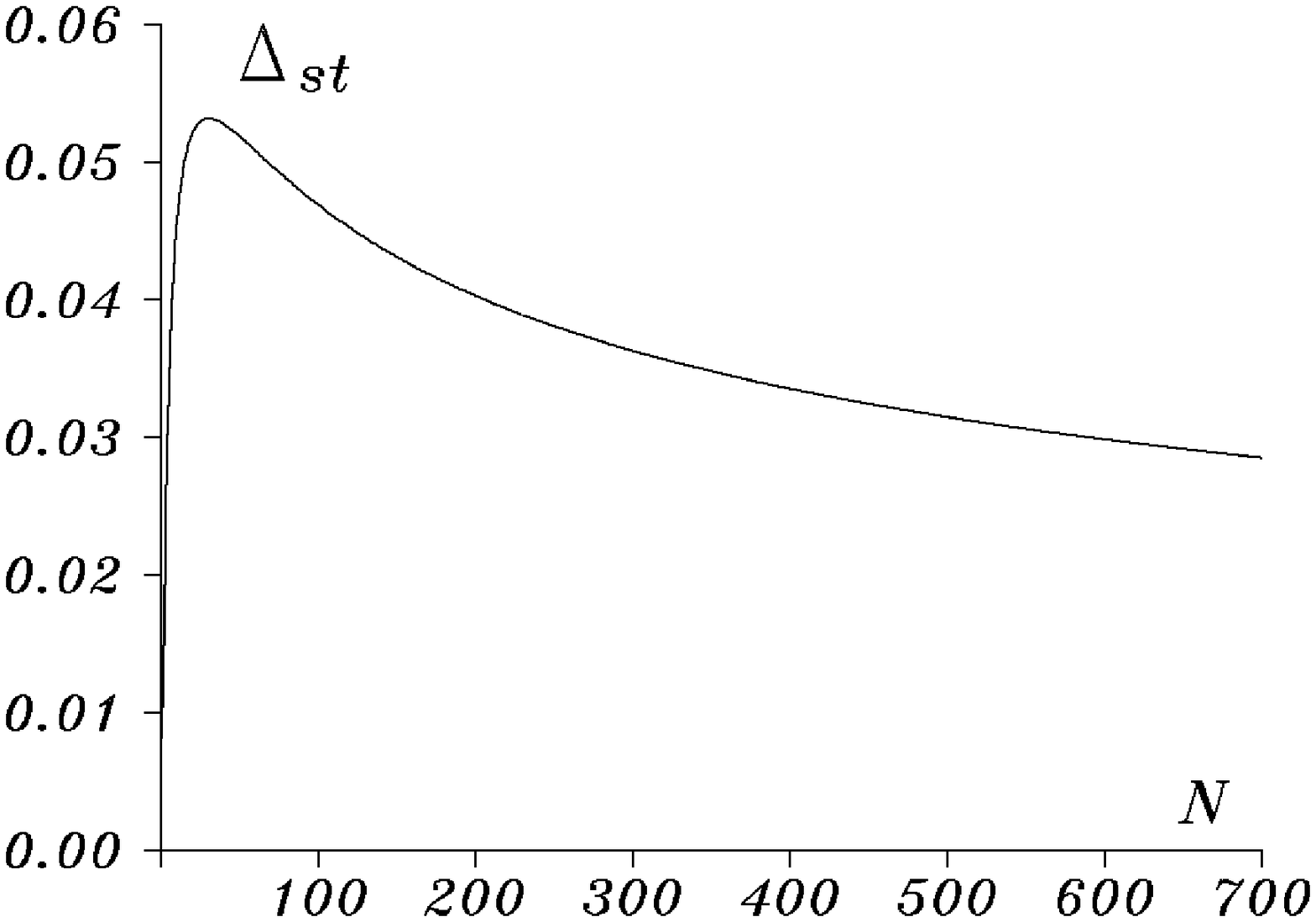}
\vglue-9.0cm \caption{Length of the stability interval as a
function of its number $N$} \label{IS}
\end{figure}

\begin{figure}
\vglue-0.0cm \hglue2.2cm
\includegraphics[width=12.0cm,keepaspectratio]{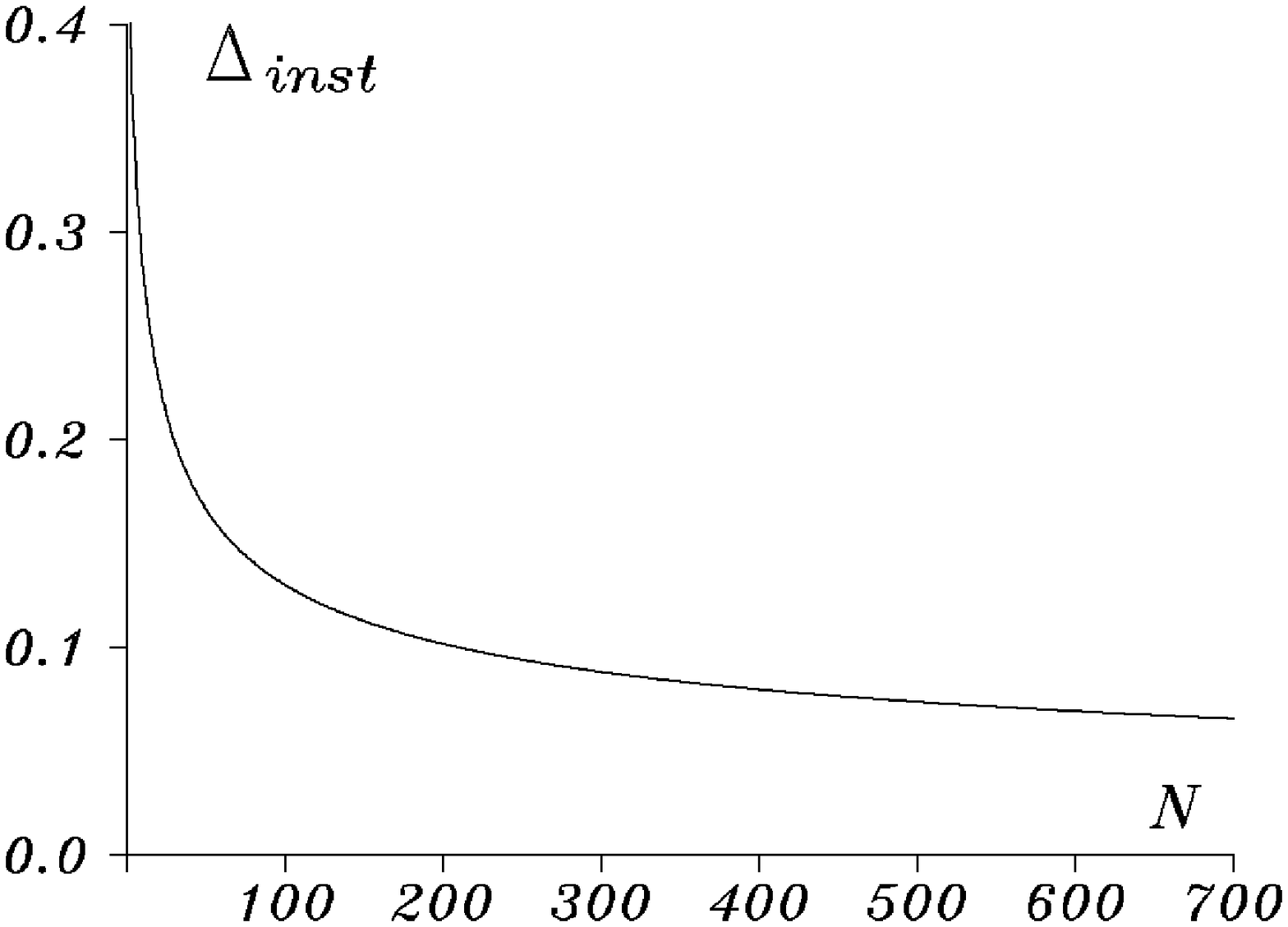}
\vglue-9.cm \caption{Length of the instability interval as a
function of its number $N$ (the first interval is not presented:
if it was shown in the same scale with all subsequent intervals,
it would have been difficult to understand the behaviour of the
graph for large $N$)} \label{II}
\end{figure}

It is also useful to discuss here in what way the length of the
stability intervals $\Delta_{st}$ and the length of the
instability intervals $\Delta_{inst}$ depend on the amplitude of
the vertical oscillations. Under the same restriction $a<a_*$ we
obtain from our numerical investigations
$$
\Delta_{st}\approx 0.3a^{-1/2},\quad \Delta_{inst}\approx
0.64a^{-1/2}.
$$

{\it Remark}. If one needs a rigorous definition about the meaning
of the quantity $a$ in the last formulae, one could interpret it
as the boundary value between two successive intervals of
stability and instability.

In Fig. 4 the behavior of the coefficients $b_1,b_2$ appearing in
the characteristic equations (10) is shown. Fig. 4a, 4b and 4c
allow us to compare the properties of these coefficients, when the
parameter $a$ varies in different intervals. All graphs
demonstrate the approximate periodicity, their period with
respect to the parameter $a$ corresponds to an increase of period
of vertical oscillations $T$ of about $8\pi$. It is important to
point out the small gaps in the Fig. 4c: for the corresponding
value of the parameter $a$ (i.e., when $a$ belongs to the
intervals where the graphs are not defined) the stability indices
$b_1,b_2$ have complex values and the so-called ``complex saddle"
instability of the vertical motion takes place. The enlarged
fragments of the graphs in the vicinity of the gaps are given in
Fig. 5.

\begin{figure}
\vglue-1.0cm \hglue2.cm
\includegraphics[width=12.0cm,keepaspectratio]{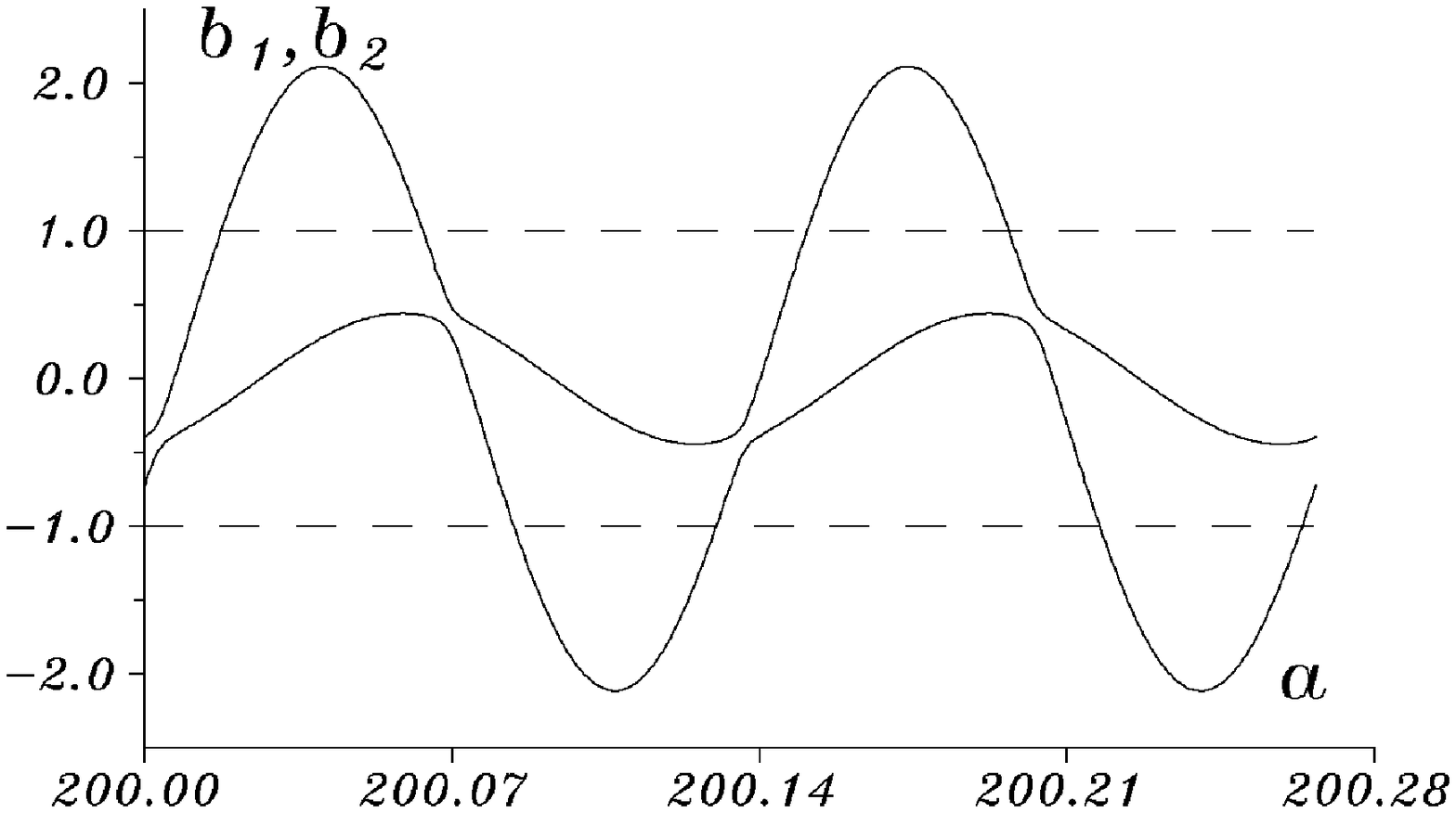}
\vglue-10.0cm \hglue2.cm
\includegraphics[width=12.0cm,keepaspectratio]{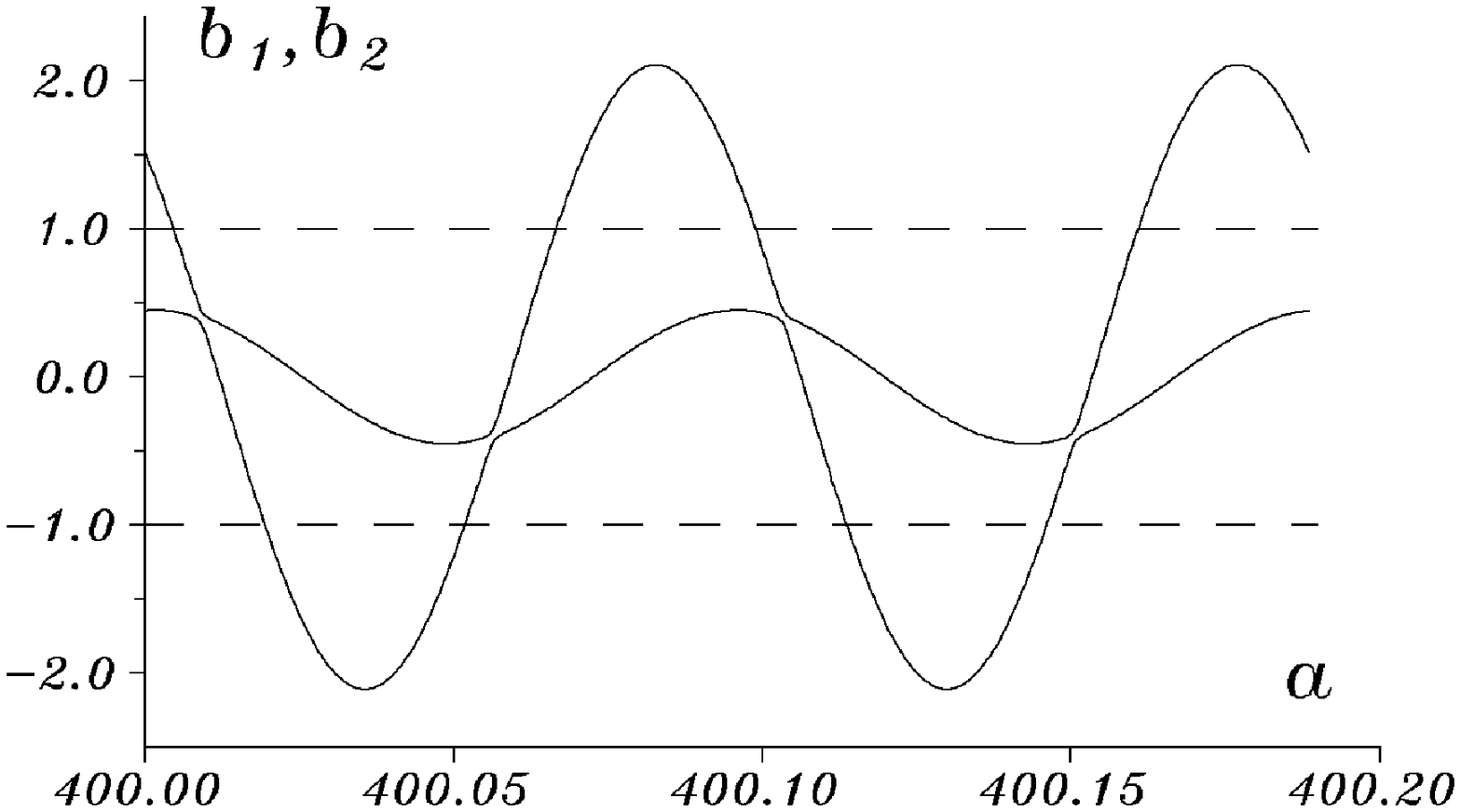}
\vglue-10.0cm \hglue2.cm
\includegraphics[width=12.0cm,keepaspectratio]{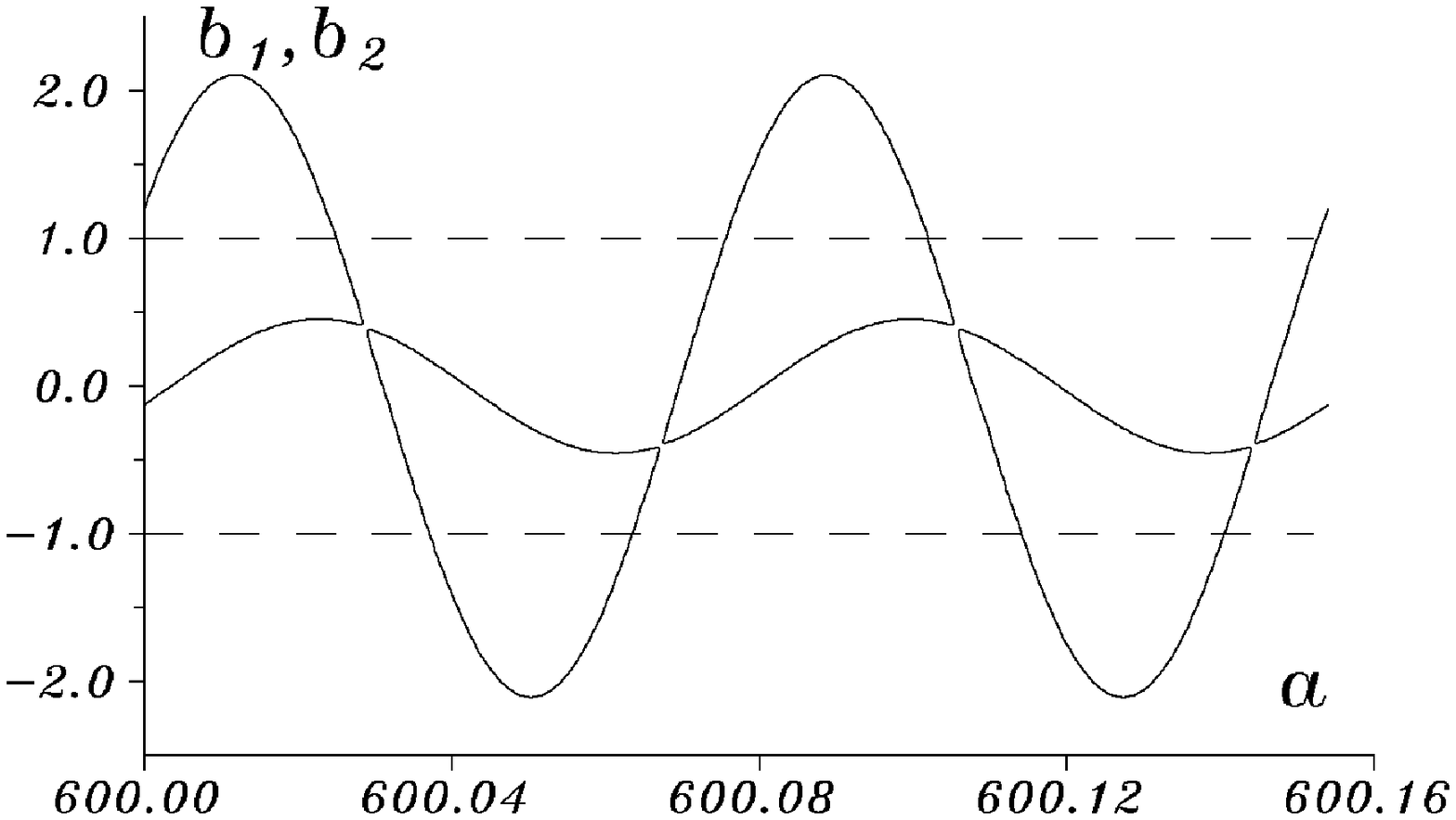}
\vglue-10.0cm \caption{The behaviour of the coefficients $b_1$ and
$b_2$ appearing in the characteristic equation (10)} \label{BB}
\end{figure}

\begin{figure}
\vglue-5.0cm \hglue2.cm
\includegraphics[width=12.0cm,keepaspectratio]{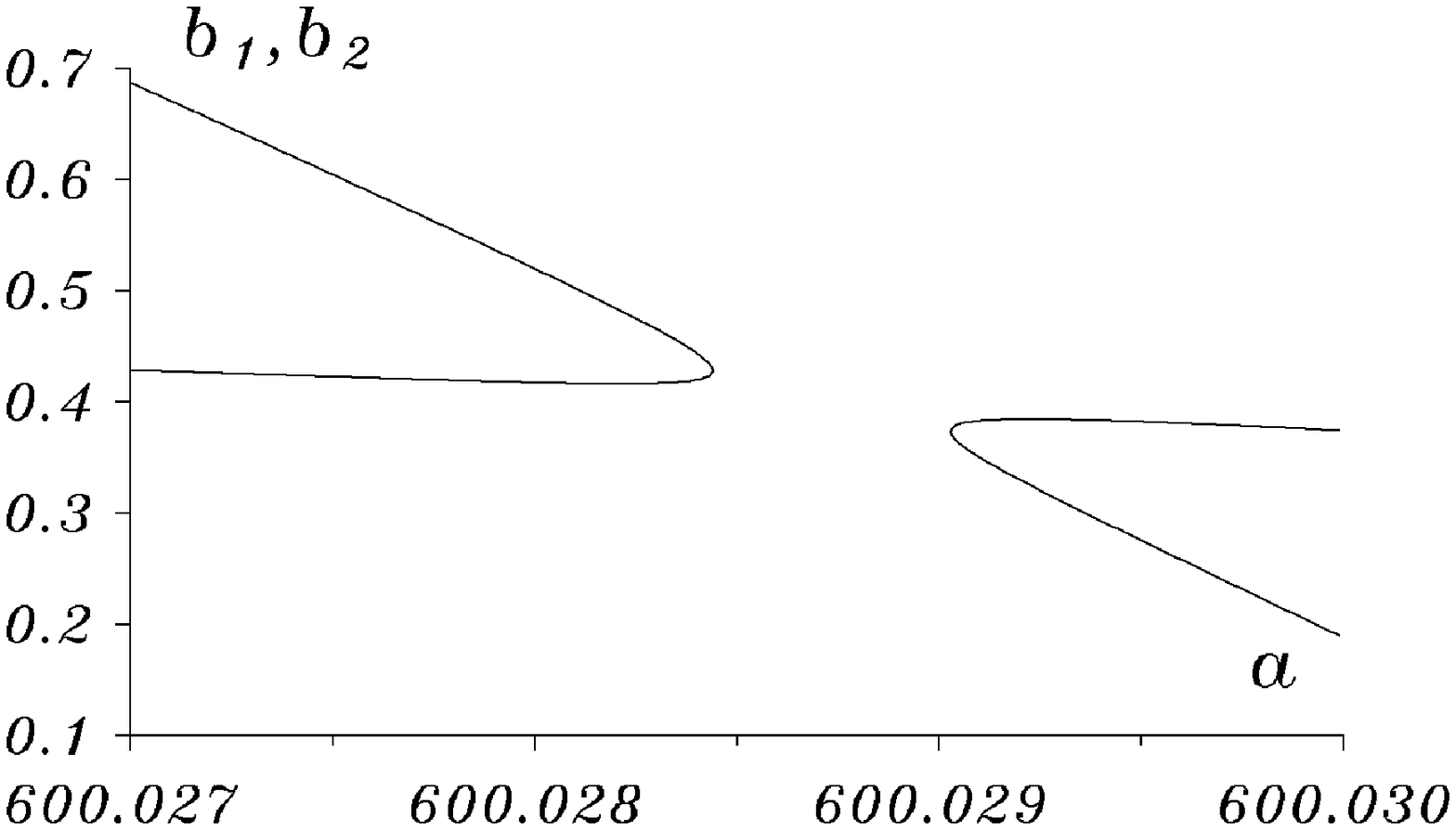}
\vglue-9.0cm \hglue1.6cm
\includegraphics[width=13.0cm,keepaspectratio]{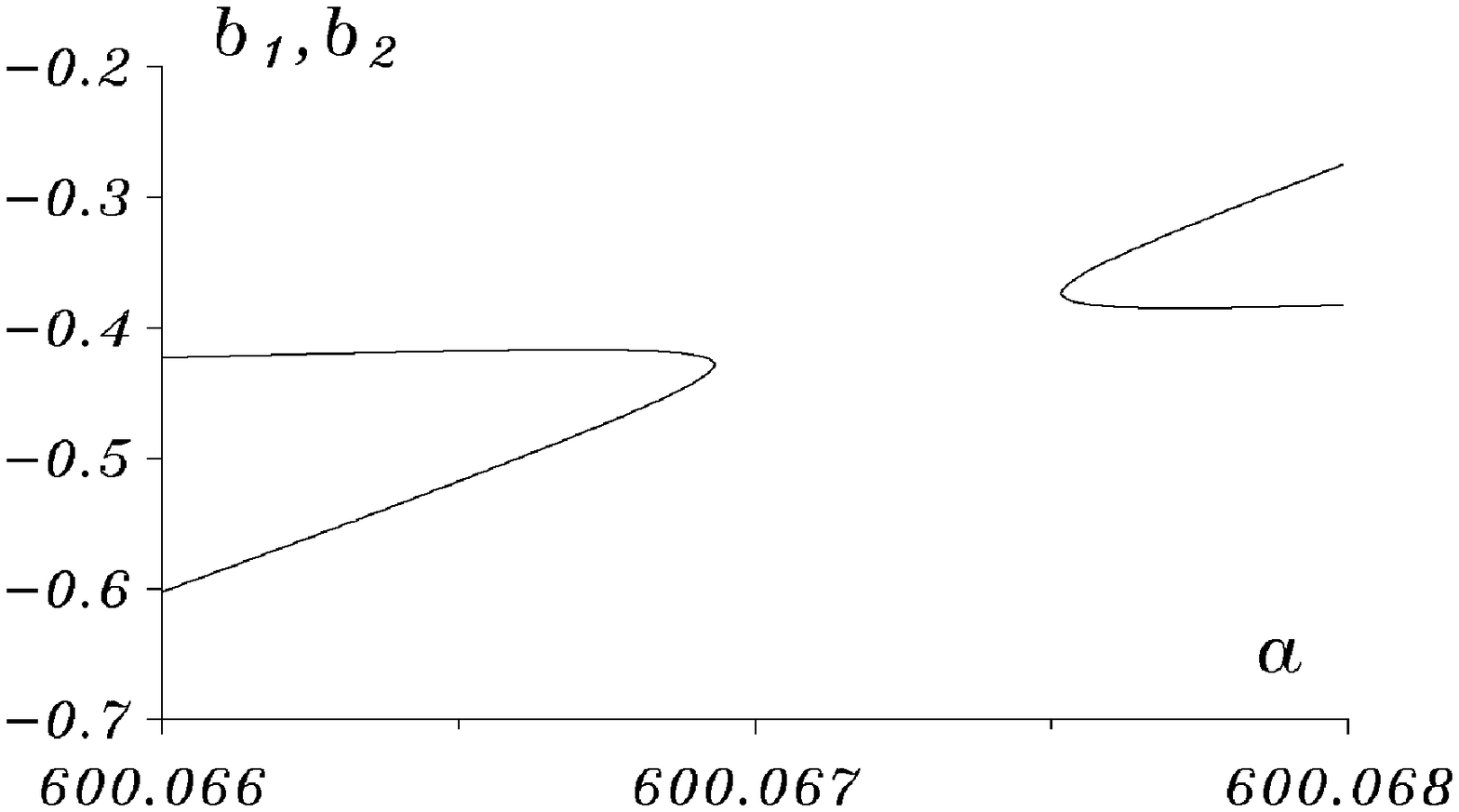}
\vglue-10.0cm \caption{The enlarged fragments of the coefficients
graphs (see Fig. 4) in the vicinity of the gaps} \label{BBZ}
\end{figure}

As it follows from our calculations, the first interval of
"complex saddle" instability begins at $a=a_*$. Since such a
value of vertical motion amplitude is large enough, it provides
us with an explanation why this kind of instability of vertical motions in
the circular Sitnikov problem was not recognized in previous
studies where relatively small values of $a$ were considered.

Increasing further the parameter $a$(i.e., for $a > a_*$), we
observe a stability/instability alternation of more complicated
type: ``wide" interval of instability - ``narrow" interval of
stability - ``narrow" interval of ``complex saddle" instability -
"wide" interval of stability - ``wide" interval of instability -
... An analog of the formulae (11) can be constructed in the case
$a>a_*$, but we prefer to present in Sect. 6 several asymptotics
written in a more convenient way.

Finally it worth while to mention  that the Runge-Kutta-Fehlberg method of 7-8
order with variable step was used to integrate numerically the variation
equations (6). The accuracy of the integration procedure (the local tolerance) was taken $10^{-10}$.
Since the period of vertical oscillations increases proportionally
$a^{3/2}$ the variation equations should be integrated over relatively large time intervals:
if we take for example $a=500$ then the value of half-period $T_* \approx 2.4837\cdot 10^4$.
To check the influence of the round-off errors some computations were
done both with double and quadruple precision arithmetic.

\section{Approximate expression for monodromy matrix}

In this section an approximate expression for the monodromy
matrix $\bM$ is derived. It will be used to discuss the phenomena
described in Sec. 4 (the alternation of stability and
instability, the decrease of stability and instability intervals
by increasing the parameter $a$, etc).

We assume that the amplitude $a$ of the periodic solution (2) is
so large, that we can define an auxiliary quantity $d$ such that
$$
1\ll d \ll a. \eqno(12)
$$

To start with we write down the monodromy matrix $\bM=\bW(T_*,0)$
as the product of three fundamental matrices:
$$
\bM=\bW(T_*,t_d^-)\bW(t_d^-,t_d^+)\bW(t_d^+,0), \eqno(13)
$$
where $t_d^+\in \left(0,\frac{T_*}{2}\right)$ and
$t_d^-=T_*-t_d^+$ are the instants at which the third body is at
distance $d$ from the barycenter $O$ in the periodic vertical
motion (2) (at $t=t_d^+$ the third body moves away from
barycenter, at $t=t_d^-$ it approaches the barycenter).

{\it Approximate expression for the matrix $\bW(t_d^+,0)$}. If the
condition (12) is satisfied the phase point $(p_3(t),q_3(t))$
moves on the manifold $\tilde{\cal V}$ in close vicinity of the
separatrix $S^+$ at $t \in [0,t_d^+]$. Within such time interval
the difference between $q_3(t,a)$ and $q_3^+(t)$ is small enough.
Neglecting this difference, we replace $q_3(t,a)$ in (6) by
$q_3^+(t)$; as a consequence, the normal matrix solution
$\bW_+(t,0)$ of the obtained system provides us the suitable
approximation for $\bW(t,0)$ at $t \in [0,t_d^+]$.

The behavior of $\bW_+(t,0)$ at $t \rightarrow +\infty$ is
described by the remarkable asymptotic formula:
$$
\bW_+(t,0)\approx \bR(t)\Lambda(q_3^+(t))\bU. \eqno(14)
$$
Here
$$
\bR(t)=\left(\begin{array}{cccc}
\hphantom{-}\cos\,t&\sin\,t&0&0\\
-\sin\,t&\cos\,t&0&0\\
0&0&\hphantom{-}\cos\,t&\sin\,t\\
0&0&-\sin\,t&\cos\,t
\end{array}\right),
\quad
$$
$$
\Lambda(q_3)= \left(\begin{array}{cccc}
\frac{\displaystyle 1}{\displaystyle q_3}&0&-\sqrt{\frac{\displaystyle 2}{\displaystyle q_3}}&0\\
0&\frac{\displaystyle 1}{\displaystyle q_3}&0&-\sqrt{\frac{\displaystyle 2}{\displaystyle q_3}}\\
\sqrt{2q_3}&0&-q_3&0\\
0&\sqrt{2q_3}&0&-q_3
\end{array}\right),
$$
$$
\bU=\left(\begin{array}{rrrr}
0.3248\ldots&0.1020\ldots&-0.4664\ldots&0.2228\ldots\\
0.1302\ldots&0.1189\ldots&0.5296\ldots&-2.0211\ldots\\
1.1175\ldots&0.1718\ldots&1.4408\ldots&0.4791\ldots\\
0.2113\ldots&0.6404\ldots&0.9414\ldots&-2.5646\ldots
\end{array}\right).
$$

The derivation of the formula (14) is based on some simple ideas.
Let us take $\dd \gg 1$ and write down $\bW_+(t,0)$ as the product
$$
\bW_+(t,0)=\bW_+(t,t_{\dd})\bW_+(t_{\dd},0), \eqno(15)
$$
where $t_{\dd}$ is the moment of time when the third body is at
distance $\dd$ from the barycenter $O$ in the motion
corresponding to the parabolic escape $q_3=q_3^+(t)$. As next
step, we modify the equations (6) to find the approximate
expression for $\bW_+(t,t_{\dd})$ at $t>t_{\dd}$. Since at
$t>t_{\dd}$ the third body is far enough from the primaries $m_1$
and $m_2$, it looks natural to replace $D$ by $q_3^+(t)$ in the
right parts of the first two equations in system (6) and to
neglect the small term $\frac{3}{4D^5}$. The system (6) takes the
form
$$
\frac{d\bz}{dt}=\bJ \overline{\bH}(q_3^+(t))\bz \eqno(16),
$$
with
$$
\overline{\bH}(q_3)=\left(\begin{array}{crcc} 1&0&0&1\\0&1&-1&0\\
0&-1&\frac{\displaystyle 1}{\displaystyle q_3^3}&0\\
1&0&0&\frac{\displaystyle 1}{\displaystyle q_3^3}\end{array}
\right).
$$

Now it is worthwhile to make the following remark. Let us
consider the rectilinear parabolic escape of the material point
in the field of an attracting center. Under a proper choice of
units, the distance between the attracting center and the point
varies as
$$
\dq(t)=\left(\frac{3}{\sqrt{2}}\right)^{2/3}t^{2/3}. \eqno(17)
$$
If the asymptotics (5) is used for $q_3^+(t)$ in the equations
(16), then these equations coincide with the motion equations of
the above mentioned material point, linearized in the vicinity of
the solution (17) and written in the reference frame uniformly
rotating around the line of the escape.

Taking this into account, we implement in (16) the change of
variables
$$
\bz=(p_1,p_2,q_1,q_2)^T \mapsto
\dbz=(\ddp_1,\ddp_2,\dq_1,\dq_2)^T,
$$
where
$$
\dbz=\bR(t_{\dd}-t)\bz.
$$
This change of variables can be interpreted as the transfer from
the synodic reference frame $Ox_1 x_2 x_3$ to the sidereal (fixed)
reference frame $O \dx_1 \dx_2 \dx_3$ ($Ox_3\|O\dx_3$). As a
result the linearized equations of motion  split into two
independent subsystems
$$
\frac{d\ddp_i}{dt}=-\frac{\dq_i}{\dq_3^3},\quad
\frac{d\dq_i}{dt}=\ddp_i,\qquad i=1,2. \eqno(18)
$$
It is not difficult to find partial solutions to the system (18)
$$
\ddp_i=\dot{\dq}_3=\sqrt{\frac{2}{\dq_3}},\quad \dq_i=\dq_3,
\quad \ddp_{3-i}\equiv 0,\quad \dq_{3-i}\equiv 0,\quad i=1,2
$$
and
$$
\ddp_i=\frac{1}{\dq_3},\quad
\dq_i=\dq_3\dot{\dq_3}=\sqrt{2\dq_3},\quad \ddp_{3-i}\equiv
0,\quad \dq_{3-i}\equiv 0,\quad i=1,2.
$$
Here and below the dots are used for derivatives with respect to
time.

Four independent partial solutions allow us to write down the
normal fundamental matrix in terms of the variables $\dbz$:
$$
\overline{\bW}_+(t,t_{\dd})=\Lambda(\dq_3(t))\Lambda^{-1}(\dq_3(t_{\dd}))
\approx \Lambda(q_3^+(t))\Lambda^{-1}(\dd).
$$
Coming back to the initial variables, we get
$$
\bW(t,t_{\dd})=\bR(t-t_{\dd})\overline{\bW}_+(t,t_{\dd}).
\eqno(19)
$$

Substituting (19) into (15) we obtain the expression for the
normal fundamental matrix $\bW_+(t,0)$ as the product of three
matrices with only one of them depending on time:
$$
\bW_+(t,0)\approx\bR(t)\Lambda(q_3^+(t))\dbU(\dd). \eqno(20)
$$
Here
$$
\dbU(\dd)=\Lambda^{-1}(\dd)\bR(-t_{\dd})\bW(t_{\dd},0).
$$
The formula (20) can be used to compute the elements of the matrix
$\bW_+(t,0)$ at $t\gg 1$. Asymptotically their values should not
depend on the choice of $\dd$. It means that the following limit
exists:
$$
\bU=\lim_{{\dd}\to+\infty}\dbU(\dd).
$$
Substituting $\bU$ instead of $\dbU(\dd)$ into (20) we arrive to
the formula (14).

The fundamental matrix $\bW_+(t,0)$ was introduced in such a way
that it provides the vertical motions satisfying (12) with a
"universal" (i.e., independent on $a$) approximation
$\bW(t,0)\approx \bW_+(t,0)$ at $t \in [0,t^+_d]$. Using the
relation (14), we finally obtain
$$
\bW(t^+_d,0)\approx \bR(t^+_d)\Lambda(d)\bU. \eqno(21)
$$

{\it Approximate expression for the matrix $\bW(t_d^-,t_d^+)$}.
Since at $t\in[t_d^+,t_d^-]$ the third body is far enough from the
primaries, we neglect again the difference between their gravity
field and the gravity field of the attracting center placed at the
barycenter $O$. To obtain the expression for $\bW(t_d^-,t_d^+)$
within such an approximation we need to integrate the system
$$
\frac{d\bz}{dt}=\bJ \overline{\bH}(\hq_3(t,a))\bz, \eqno(22)
$$
where $\hq_3(t,a)$ describes the motion in the Newtonian field
along the segment $[0,a]$ on the axis $Ox_3$. It is supposed that
the maximum distance $a$ from the body to the attracting center is
achieved at $t=\frac{T_*}{2}$. In this case $q_3(t,a)\approx
\hq_3(t,a)$ at $t\in[t_d^+,t_d^-]$. Of course the motion along a
segment corresponds to the singular impact orbit~\cite{S1967},
but it is used here to approximate the regular vertical motion on
the time interval were the singularities are absent.

The change of variables
$$
\bz=(p_1,p_2,q_1,q_2)^T \mapsto \hbz=(\hp_1,\hp_2,\hq_1,\hq_2)^T,
$$
where
$$
\hbz=\bR\left(\frac{T_*}{2}-t\right)\bz, \eqno(23)
$$
allows us to rewrite the equations (22) in the more simple form:
$$
\frac{d\hp_i}{dt}=-\frac{\hq_i}{\hq_3^3},\quad
\frac{d\hq_i}{dt}=\hp_i,\qquad i=1,2. \eqno(24)
$$
It is easy to check that the system (24) admits the following
partial solutions:
$$
\hp_i=\dot{\hq}_3,\quad \hq_i=\hq_3, \quad \hp_{3-i}\equiv
0,\quad \hq_{3-i}\equiv 0,\quad i=1,2 \eqno(25)
$$
and
$$
\hp_i=\frac{1}{\hq_3}-\frac{2}{a},\quad
\hq_i=\hq_3\dot{\hq_3},\quad \ddp_{3-i}\equiv 0,\quad
\dq_{3-i}\equiv 0,\quad i=1,2. \eqno(26)
$$

To compute $\dot{\hq}_3$ in (25) and (26) the energy integral can
be used. In the case of the motion along the segment $[0,a]$ in
the Newtonian field, this integral takes the form
$$
\frac{\dot{\hq}_3^2}{2}-\frac{1}{\hq_3}=-\frac{1}{a},
$$
and consequently
$$
\dot{\hq}_3(t)=\pm
\sqrt{2\left(\frac{1}{\hq_3(t)}-\frac{1}{a}\right)}.
$$

Taking into account (25),(26) we write down the fundamental matrix
for the system (24) as
$$
\hat{\bW}\left(t,\frac{T_*}{2}\right)= \left(\begin{array}{cccc}
2-\frac{\displaystyle a}{\displaystyle \hq_3}&0&\frac{\displaystyle \dot{\hq}_3}{\displaystyle a}&0\\
0&2-\frac{\displaystyle a}{\displaystyle \hq_3}&0&\frac{\displaystyle \dot{\hq}_3}{\displaystyle a}\\
-a\hq_3\dot{\hq}_3&0&\frac{\displaystyle \hq_3}{\displaystyle
a}&0\\
0&-a\hq_3\dot{\hq}_3&0&\frac{\displaystyle \hq_3}{\displaystyle a}
\end{array}\right)
$$
and then (taking into account the relation (23)) we write the
matrix for the system (22)
$$
\bW\left(t,\frac{T_*}{2}\right)=\bR\left(t-\frac{T_*}{2}\right)
\hat{\bW}\left(t,\frac{T_*}{2}\right). \eqno(27)
$$

Using the expression (27) we find
$$
\bW\left(t_d^-,\frac{T_*}{2}\right)=\bR\left(t_d^-
-\frac{T_*}{2}\right)\bN(d,a),
$$
where we denote by
$$
\bN(d,a)=\hat{\bW}\left(t_d^-,\frac{T_*}{2}\right)=
$$
$$
\left(\begin{array}{cccc}
2-\frac{\displaystyle a}{\displaystyle
d}&0&-\frac{\displaystyle 1}{\displaystyle
a}\sqrt{2\left(\frac{\displaystyle 1}{\displaystyle
d}-\frac{\displaystyle 1}{\displaystyle a}\right)}&0\\
0&2-\frac{\displaystyle a}{\displaystyle d}&0&-\frac{\displaystyle
1}{\displaystyle a}\sqrt{2\left(\frac{\displaystyle
1}{\displaystyle
d}-\frac{\displaystyle 1}{\displaystyle a}\right)}\\
d\sqrt{2\left(\frac{\displaystyle a}{\displaystyle
d}-1\right)}&0&\frac{\displaystyle d}{\displaystyle a}&0\\
0&d\sqrt{2\left(\frac{\displaystyle a}{\displaystyle
d}-1\right)}&0&\frac{\displaystyle d}{\displaystyle a}
\end{array}\right).
$$

The final step is based on the last formula in (8), namely
$$
\bW\left(t_d^-,t_d^+\right)=\bW\left(t_d^-,\frac{T_*}{2}\right)\bW\left(\frac{T_*}{2},t_d^+\right)=
$$
$$
\bW\left(t_d^-,\frac{T_*}{2}\right)\bQ\bW^{-1}\left(t_d^-,\frac{T_*}{2}\right)\approx
\bR(t_d^+-t_d^-)\bK(d). \eqno(28)
$$
Here
$$
\bK(d)=\left(\begin{array}{cccc}
-3&0&2\sqrt{2}d^{-3/2}&0\\
0&-3&0&2\sqrt{2}d^{-3/2}\\
2\sqrt{2}d^{3/2}&0&-3&0\\
0&2\sqrt{2}d^{3/2}&0&-3 \end{array}\right)
$$
For completeness we should add the following formula:
$$
\bK(d)\approx \bN(d,a)\bQ\bN^{-1}(d,a)\bQ.
$$

{\it Approximate expression for the matrix $\bW(T_*,t_d^-)$}.
Using again the relations (8) we get
$$
\bW(T_*,t_d^-)=\bW(0,-t_d^+)=\bQ\bW^{-1}(t_d^+,0)\bQ. \eqno(29)
$$
Then the substitution of the previously obtained expression for
$\bW(t_d^+,0)$ (the formula (21)) into the right part of (29)
provides us with the desired approximate formula for
$\bW(T_*,t_d^-)$.

{\it Finalizing the construction of the approximate formula for
the monodromy matrix $\bM$}. The substitution of the approximate
expressions for $\bW(T_*,t_d^-)$, $\bW(t_d^-,t_d^+)$,
$\bW(t_d^+,0)$ into (13) yields
$$
\bM(a)\approx
\left\{\bQ\left(\Lambda(d)\bU\right)^{-1}\bQ\bR(T_*-t_d^-)\right\}
\left\{\bR(t_d^--t_d^+)\bK(d)\right\}
\left\{\bR(t_d^+)\Lambda(d)\bU\right\}. \eqno(30)
$$
Simplifying (30) the extraordinary simple result can be obtained
$$
\bM(a)\approx\bQ\bU^{-1}\bQ\bR(T_*(a))\bU. \eqno(31)
$$
This formula allow us to investigate analytically the stability
properties of the periodic vertical motions in the case $a \gg 1$.

\section{New insight into the stability properties of the vertical
motions}

As it follows from (31) the coefficients of the monodromy matrix
$\bM$ and, respectively, the coefficients of the characteristic
equation (9) are $2\pi$-periodic functions of the semiperiod of
the vertical motion $T_*$. To illustrate this we present in Fig.6
the graphs of the coefficients $b_1,b_2$ (only the real values)
as $T_*$ varies in the interval $[2\pi n,2\pi(n+1)]$, where $n$
is a large enough integer number. Using the formula (3), which
defines the dependence of $T_*$ on the amplitude $a$, it is not
difficult to prove that in terms of $a$ the lengths of the
stability and instability intervals decrease proportionally to
$a^{-1/2}$ as $a\to +\infty$.

\begin{figure}
\vglue0cm \hglue1.cm
\includegraphics[width=14.0cm,keepaspectratio]{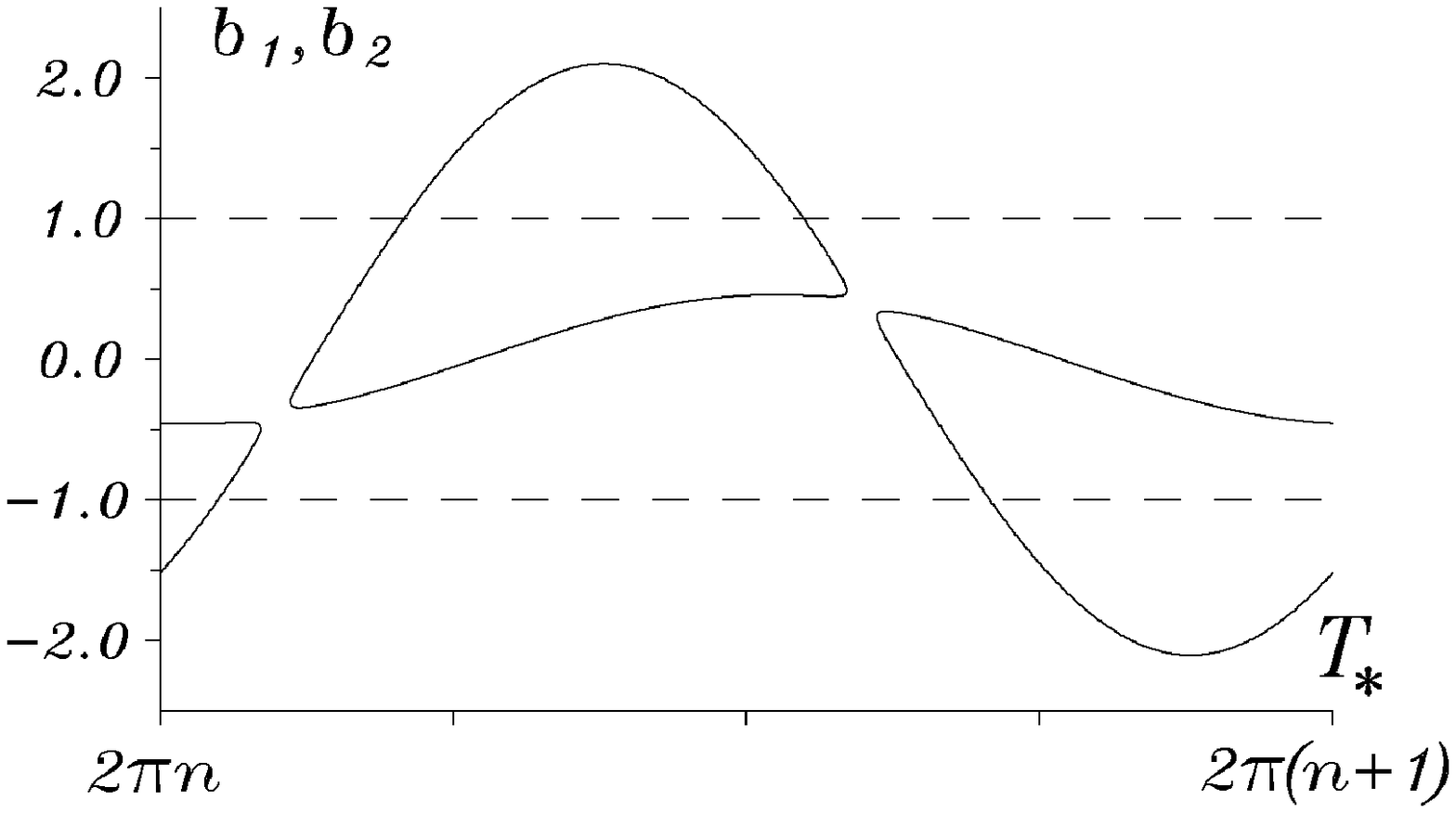}
\vglue-13cm \caption{Graphs of the coefficients $b_1,b_2$
computed on the base of the approximate formula for the monodromy
matrix $\bM$. Only real values are shown.} \label{L}
\end{figure}

Taking into account the approximate expression for the monodromy
matrix $\bM$, we describe in more details the repeating pattern
of stable and unstable intervals mentioned at the end of Sec. 4.
This pattern consists of four intervals appearing in the
following order as $a$ increases:

{\it ``Wide" interval of instability.} Both coefficients $b_1,b_2$
are real, but one of them has absolute value greater than 1
("saddle-center" instability). The asymptotic length of the
interval in terms of the amplitude of the motion  is about
$0.643544\cdot a^{-1/2}$, while the variation of the semiperiod
$T_*$ equals to about $2.144392$.

{\it ``Narrow" interval of stability.} The coefficients $b_1,b_2$
are real and belong to the interval $(-1,1)$. The approximate
length is $0.068655\cdot a^{-1/2}$; the variation of the
semiperiod equals to about $0.228768$.

{\it ``Narrow" interval of instability.} The coefficients
$b_1,b_2$ are complex ("complex saddle" instability). The
approximate length is $0.048166\cdot a^{-1/2}$; the variation of
the semiperiod equals to about $0.160497$.

{\it ``Wide" interval of stability.} The coefficients $b_1,b_2$
are real and belong to the interval $(-1,1)$ again. The
approximate length is $0.182445\cdot a^{-1/2}$; the variation of
the semiperiod equals to about $0.607936$.

To conclude, we recall that before the first appearance of the
interval of ``complex saddle" instability at $a=a_*$, a more
simple pattern with only two intervals was observed. The
"transient" asymptotics for the length of the stability intervals
in the case $1\ll a < a_*$ can be obtained by adding of the
lengths of the ``narrow" instability interval and both stability
intervals in the final pattern. It yields
$$
\Delta^{tr}_{st}\approx 0.299\cdot a^{-1/2},
$$
which is in good agreement with the corresponding numerical
result presented in Sec. 4.

\section{Stability of the vertical motions in the circular Sitnikov
problem with four and more bodies}

The investigation of the generalized circular Sitnikov problem
with four and more bodies revealed that in contrast to the case of
the three body problem there is no alternation of
stability/instability in the family of vertical
motions~\cite{BP2009,SPB2008}.

For simplicity we limit our consideration to the case of the
restricted four body problem. It is assumed that three primaries
of equal mass rotate around the barycenter $O$ in circular orbit
with the radius $R=1/\sqrt{3}$~\cite{SPB2008}. Under the linear
approximation the stability analysis of the fourth body periodic
vertical motion $\bq(t,a)$ is reduced to the study of the
spectral properties of the monodromy matrix associated to the
system of linear differential equations with periodic coefficients
$$
\frac{d\bz}{dt}=\bJ\bbH(t)\bz, \eqno(32)
$$
where
$$
\bbH(t)=\left(\begin{array}{cccc} 1&0&0&1\\
0 & 1 & -1 & 0 \\
0 & -1 & \left(\frac{\displaystyle 1}{\displaystyle D^3}-
\frac{\displaystyle 1}{\displaystyle 2D^5}\right)& 0 \\
1 & 0 & 0 &\left(\frac{\displaystyle 1}{\displaystyle D^3}-
\frac{\displaystyle 1}{\displaystyle 2D^5}\right)\end{array}
\right),
$$
$$
D(t,a)=\left(\bq_3^2(t,a)+\frac{1}{3}\right)^{1/2}.
$$

It is remarkable that equations (32) possess a circular symmetry:
for any real $\alpha$ they are invariant with respect to
transformations of the form
$$
\widetilde{\bz}=\bR(\alpha)\bz,
$$
while the non-linearized equations of motion  of the fourth body
in the synodic reference frame admit only the rotational symmetry
of the 3rd order. The possibility for the linearized equations of
motion  to have a larger group of symmetries in comparison to the
original non-linear system was pointed out by V.I.
Arnold~\cite{A1989}(Sec. 23). In particular, in the case of the
initial system rotational symmetry of $N$-th order ($N\ge 3$) the
linearized equations always have circular symmetry. This is the
reason why the stability analysis of the vertical motions, based
on the linearized equations, yields similar results for the
Sitnikov problem with four and more bodies and for the particle
dynamics in the gravity field of the circular ring~\cite{BE2005}
(numerically it was shown in~\cite{BP2009}).

It is convenient to rewrite the equations of motion (32) in a
sidereal (fixed) reference frame by means of the transformation
of variables
$$
\bz=(p_1,p_2,q_1,q_2)^T \mapsto \bbz=(\bp_1,\bp_2,\bq_1,\bq_2)^T,
$$
where
$$
\bbz=\bR(-t)\bz.
$$
After that the equations of motion  split into two identical
independent subsystems:
$$
\frac{d\bp_i}{dt}=-\frac{\bq_i}{2D^3}\left(2-\frac{1}{D^2}\right),\quad
\frac{d\bq_i}{dt}=\bp_i, \quad i=1,2. \eqno(33)
$$

Let $\bbW_+(t,t')$ denote the normal fundamental matrix for the
system (33) in the case when $\bq_3(t,a)$ is replaced by
$\bq_3^+(t)$, which corresponds to the parabolic escape. Using
the same technique as in Sect. 5 we obtain the asymptotic formula
$$
\bbW_+(t,0)\approx \Lambda(\bq_3^+(t))\bbU,
$$
where
$$
\bbU=\lim_{d\to+\infty}\Lambda^{-1}(d)\bbW_+(t_d,0)\approx
\left(\begin{array}{cccc}
0.2456&0&-1.2690&0\\
0&0.2456&0&-1.2690\\
0.9246&0&-0.7061&0\\
0&0.9246&0&-0.7061
\end{array}\right).
$$

Applying the main ideas of Sect. 5, we establish the following
property of the monodromy matrix $\bbM(a)$ associated to (33): at
$a\to+\infty$ the matrix $\bbM(a)\to \bbM_*$, where the constant
matrix $\bbM_*=\bQ\bbU^{-1}\bQ\bbU$. The eigenvalues of the matrix
$\bbM_*$ are the asymptotic limits for multiplicators (of
multiplicity $2$) of the system (33):
$$
\lim_{a\to+\infty}\brho_i(a)=\brho_i^*,\quad
\brho_1^*=-0.4446\ldots,\,\brho_2^*=-2.2488\ldots
$$

Finally, it is not difficult to derive the asymptotic formulae the
for multiplicators of the original system (32):
$$
\rho_{1,2}\approx\brho_1^*\exp(\pm iT_*),\quad
\rho_{3,4}\approx\brho_2^*\exp(\pm iT_*).
$$
On the complex plane $\rho_1,\ldots,\rho_4$ are placed in the
small vicinity of the circles with radii $|\rho_1|<1$ and
$|\rho_2|>1$. Consequently in the circular Sitnikov problem with
four bodies the periodic vertical motions  with large amplitudes
are always unstable.

Finally we would like to note the another opportunity to introduce
the generalized circular Sitnikov problem with $N$ bodies using
the appropriate straight line solution of the problem of $(N-1)$
bodies~\cite{M1910}. If in such a solution $(N-1)$ primaries are
arranged symmetrically with respect to the barycenter then the infinitesimal
$N$th body can move periodically along an axis around which the rotation of the
primaries takes place (naturally, the proposed generalization is possible
for odd $N$ only). Likely this family of periodic motions exhibits the alternation
of stability and instability.

\section{Conclusion}

The combination of numerical and analytical approaches provided
us with the opportunity to correct, clarify and extend some
previously known results related to the circular Sitnikov problem
(mainly about the stability of vertical motions). For the first
time under the scope of this problem the possibility of the
"complex saddle" instability was revealed within the family of
vertical motions.

For our theoretical constructions it was essential that the phase
trajectories corresponding to the solution under consideration
have lengthy parts in the vicinity of the peculiar separatrices of
the problem - the parabolic escapes to infinity. Often enough it
is possible to introduce a suitable auxiliary mapping in the
vicinity of the separatrix in order to study the local properties
of the phase flow. It would be very interesting to develop
similar for the circular Sitnikov problem.\vskip.1in

\noindent {\bf Acknowledgements.} The author would like to
express his gratitude to A.I. Neishtadt and A.B. Batkhin for
useful discussions during the accomplishment of this work. Also
the author thanks A.Celletti for reading the manuscript and
suggesting many improvements.

This work was partially supported from the
Russian foundation for Basic Research via Grant NSh-6700.2010.1.

\end{document}